# Instrument for the assessment of road user automated vehicle acceptance:
# A pyramid of user needs of automated vehicles

Sina Nordhoff [a], Marjan Hagenzieker [a], Esko Lehtonen [b], Michael Oehl [c], Marc Wilbrink [c], Ibrahim Ozturk [d], David Maggi [e], Natacha Métayer [f], Gaëtan Merlhiot [f], Natasha Merat [d]

[a] Department Transport & Planning, Delft University of Technology, The Netherlands
[b] VTT Technical Research Centre of Finland Ltd., P.O. Box 1000, FI-02044 Espoo, Finland
[c] German Aerospace Center, Koeln, Germany
[d] Institute for Transport Studies, University of Leeds, Leeds LS2 9JT, UK
[e] Volvo Group Trucks Technology, Gothenburg, Sweden
[f] VEDECOM Institute, MobiLAB, 23 bis Allée des Marronniers, 78000, Versailles, France

## ABSTRACT

This study proposed a new methodological approach for the assessment of automated vehicle acceptance (AVA) from the perspective of road users inside and outside of AVs pre- and post- AV experience. Users can be drivers and passengers, but also external road users, such as pedestrians, (motor-)cyclists, and other car drivers, interacting with AVs. A pyramid was developed, which provides a hierarchical representation of user needs. Fundamental user needs are organized at the bottom of the pyramid, while higher-level user needs are at the top of the pyramid. The pyramid distinguishes between six levels of needs, which are safety trust, efficiency, comfort and pleasure, social influence, and well-being. Some user needs universally exist across users, while some are user-specific needs. These needs are translated into operationalizable indicators representing items of a questionnaire for the assessment of AVA of users inside and outside AVs. The formulation of the questionnaire items was derived from established technology acceptance models. As the instrument was based on the same model for all road users, the comparison of AVA between different road users is now possible. We recommend future research to validate this questionnaire, administering it in studies to contribute to the development of a short, efficient, and standardized metric for the assessment of AVA.

## 1.    Introduction

The field of automated vehicle acceptance (AVA) has gained enormous interest in the past few years. Establishing user acceptance of automated vehicles (AVs) is of utmost importance because if AVs are not accepted, the safety, efficiency, and equity benefits of road automation will not be realised, and the large investments in this technology will not materialize (Nordhoff, Van Arem, & Happee, 2016; Van Der Laan, Heino, & De Waard, 1997).

AVA is a multi-user phenomenon. It covers drivers of automated passenger vehicles, truck drivers, passengers, safety drivers, or external road users, such as pedestrians, (motor-)cyclists, and other car drivers, communicating and interacting with AVs on public roads (Kaye, Li, Oviedo-Trespalacios, & Pooyan Afghari, 2022; Merat & Lee, 2012). Vulnerable road users, such as pedestrians and (motor-)cyclists, have been disproportionally involved in fatal accidents (WHO, 2022). For this reason, it is important to consider the (safety) needs and preferences of not only users inside but also users outside AVs. In line with Maslow and Lewis (1987), the present paper argues that user groups share some fundamental and basic needs towards AVs, such as the need for safety, efficiency, and comfort. However, each user group also has unique needs. For example, passengers in AVs may be prone to motion sickness, while this aspect may be less relevant for other external user groups.

The field of AVA has flourished from the application of technology acceptance models in recent years. However, several limitations can be identified, which provide important imperatives for this work.

First, previous studies have mainly investigated AVA of AV users in isolation, with a main focus on drivers of automated passenger vehicles or passengers of automated shuttles (Kaye, Lewis, Forward, & Delhomme, 2020; Madigan, Louw, Wilbrink, Schieben, & Merat, 2017). Recently, studies have started to investigate the acceptance of passengers of automated passenger vehicles (Pascale et al., 2021). In addition, other works on external road users integrate some dimensions of UTAUT1/2 into their studies (Deb et al., 2017; Koustanaï et al., 2022). Deb et al. (2017) have shown that safety and interactions are key factors influencing the willingness of pedestrians to cross in front of an AV. Koustanaï et al. (2022) has shown that trust had a direct effect on the behavioral intention to share the road with an AV, while perceived behavioral control, reliability, perceived safety, attitudes and experience had indirect effects on behavioral intention.

Second, a common way to investigate technology acceptance has been the use of technology acceptance models, such as the Unified Theory of Acceptance and Use of Technology (UTAUT1/2) (Venkatesh, Morris, Davis, & Davis, 2003; Venkatesh, Thong, & Xu, 2012), which is a synthesis of eight influential technology acceptance models, including the Technology Acceptance Model (TAM) (Davis, 1985). These models posit that the behavioral intention to use technology is directly influenced by cognitive domain-specific and emotional-affective factors. Cognitive domain-specific factors include the perceived usefulness (i.e., performance expectancy), ease of use (i.e., effort expectancy), and conditions supporting the use of AVs (i.e., facilitating conditions). Emotional affective components include the perceived enjoyment of AVs (i.e., hedonic motivation), and the support of the use of AVs in the individual's social networks (Venkatesh et al., 2012). These models were specifically developed for the investigation of technology acceptance in general. When they are administered in studies investigating technology acceptance, the factors of these models are translated into measurable or operationalizable questionnaire items distributed to and rated by respondents. The wording of the items has to be adjusted to the context of road vehicle automation every time researchers aim to implement the models in their studies. The items have not been translated into measurable items for multiple road users being pivotal for AVA. Validity issues may be the result, i.e., to what extent can researchers warrant that they measure what they intend to measure when the translation to the specific research context deviates to a large extent from the original meaning? To compare AVA between different road users, we need instruments based on the same model for all the road users.

Third, another limitation of common technology acceptance models is that they do not theorize relationships between perceived safety and trust and behavioral intention, respectively. Perceived safety and trust are pivotal for AVA for all road users as they have to put their lives into the hands of a robot. If they don't feel safe and trust the AV, they may be less likely to accept and interact with them.

Fourth, no standardized instrument for measuring AVA before and after the experience with AVs exists. AVA can be measured before and after experience with AVs. Schade and Schlag (2003) defined acceptance as the *"respondents' attitudes, including their behavioral responses, after the introduction of a measure, and acceptability as the prospective judgement before such future introduction"*. According to this definition, the term 'acceptance' is applied when respondents had actual experience with AVs, where acceptability is assessed prior to experience with automated vehicles. Typically, researchers investigating AVA applied the term ‚acceptance' for research studies surveying respondents with and without physical experience with AVs. Another definition of acceptance is proposed by Adell (2010) who defined acceptance as the „*degree to which an individual intends to use a system and, when available, incorporates his system in his / her driving*" (p. 477).

## 1.1.    Research objectives

The main objective of the present paper is to develop a standardized instrument for the assessment of AVA of road users pre- and post- AV experience. The instrument consists of a standard part that can be implemented in studies across user groups. It also consists of a variable part accounting for the unique needs of each user group. The user needs were derived from the literature, and organized in a pyramid, which serves as hierarchical representation of these user needs.

## 2. Literature review

### 2.1. Safety

One of the most proclaimed AV benefits pertains to safety: AVs are expected to improve traffic safety (Pyrialakou, Gkartzonikas, Gatlin, & Gkritza, 2020). Safety has an objective and subjective dimension (Nilsen et al., 2004). The objective dimension of AV safety has been commonly investigated in simulation studies by the number of AV crashes in relation to mileage and safety-critical AV behavior (Kalra & Paddock, 2016). The safety of AVs is particularly important for truck drivers and fleet owners who are ultimately responsible for third-party goods (Othman, 2021). The subjective dimension captures the individual's subjective feelings (Nordhoff, Stapel, He, Gentner, & Happee, 2021; Xu et al., 2018). Recently, the attention of scientific scholars has shifted from the objective dimension to the consideration of perceived safety for AVA. Studies mainly investigated the perceived safety of users inside rather than users outside AVs (Pammer, Gauld, McKerral, & Reeves, 2021; Parkin et al., 2022; Pyrialakou et al., 2020; Vlakveld, van der Kint, & Hagenzieker, 2020). In our previous study with users of partially automated cars (Nordhoff et al., 2021), perceived safety influenced automation use indirectly through trust. In other studies, perceived safety did have a direct impact on the intention to use AVs (Montoro et al., 2019; Xu et al., 2018). Whether external road users are and feel safe around AVs will depend to a large extent on how they communicate and interact with AVs. External road users can communicate via internal and external communication means with AVs. Pedestrians reported to rely on vehicle kinematics, such as vehicle speed or gap distance, to inform their decision to cross the road in front of AVs (Wang et al., 2021). To address the lack of hand gestures and eye contact by human drivers in driverless vehicles, external Human Machine Interfaces (eHMIs) as external communication displays located on the outside of AVs indicating vehicle intent have been proposed. Studies currently count to around 70 eHMI concepts, which were mainly designed from the perspective of pedestrians (Berge, Hagenzieker, Farah, & de Winter, 2022; Dey et al., 2020). It is unclear whether eHMIs serve as 'gimmicks' or 'necessity' for enabling safe interactions between AVs and external road users (de Winter & Dodou, 2022). External road users, such as pedestrians and cyclists, may rely more on implicit communication forms, such as vehicle kinematics, rather than eye contact or body gestures in their interactions with human drivers (de Winter & Dodou, 2022; Fridman et al., 2017). To inform their crossing decisions in front of an AV, (**Error! Bookmark not defined.**motor-)cyclists preferred to receive instructions from the AV (e.g., go ahead) to the status of the AV (Pammer et al., 2021). Currently, driverless automated shuttle services (e.g., Waymo, Cruise) operate on public roads in e.g., San Francisco, without any external communication interfaces. It is plausible that external communications interfaces are not needed to enable safe and acceptable interactions with external road users. A study with cyclists / motorcyclists conducted by Pammer et al. (2021) revealed that respondents expected 'fewer crashes' and 'reduced severity of crashes' to be a perceived benefit of AVs. In this study, cycling near an AV was considered the least unsafe scenario, followed by walking and driving near an AV. Xing, Zhou, Han, Zhang, and Lu (2022) observed that vulnerable road users had more positive perceptions of AV safety in 2019 rather than 2017 (increase by around 10% to 30%). In the study of Berge et al. (2022), respondents mentioned the potential of on-bike eHMIs to increase the safety of cyclists. We posit here that (perceived) safety is a fundamental human need at the bottom of the pyramid as shown by Figure 1, which can be translated into measurable indicators for safety for all road users ("arrive more safely", "feel safer") (see Table 1).

## 2.2. Trust

Trust in technology has been considered a fundamental factor impacting how humans interact with technology (Lee & See, 2004). Previous studies supported the role of trust as positive predictor of the behavioral intention to use AVs (Benleulmi & Ramdani, 2022; Du, Zhu, & Zheng, 2021; Foroughi et al., 2023; Kaur & Rampersad, 2018; Kenesei et al., 2022; Kettles & Van Belle, 2019; Meyer-Waarden & Cloarec, 2022; Panagiotopoulos & Dimitrakopoulos, 2018; Waung, McAuslan, & Lakshmanan, 2021; Xu et al., 2018; Zhang et al., 2020). Reliability of automation is a key factor impacting trust, with an increase in reliability contributing to an increase in trust (Carsten & Martens, 2019). Drivers failing to monitor automation (i.e., complacency) has been associated with overtrust in automation (Banks, Eriksson, O'Donoghue, & Stanton, 2018; Nordhoff et al., 2023; Wilson, Yang, Roady, Kuo, & Lenné, 2020). Failing to monitor automation is not only a concern for users inside AVs: Research indicates that pedestrians intentionally stepped in front of an AV to test its capabilities and limitations (Madigan et al., 2019), reported an intention to bully AVs (Liu, Du, Wang, & Da Young, 2020), or showed other types of aggressive behaviors towards AVs (Haué, Merlhiot, Koustanaï, Barré, & Moneger), such as choosing shorter gap distances in comparison to conventional vehicles (Dommes et al., 2021). Scholars observed that cyclists / motorcyclists had higher trust in human drivers than general trust in AVs, but reported a higher trust in AVs rather than human drivers to have their own personal safety as a priority (Pammer et al., 2021). In the study of Hagenzieker et al. (2020), cyclists indicated to have more confidence in human-driven than automated cars. They were more confident of being noticed by the AV rather than traditional car when they had priority over the car, while they were more confident of being noticed by the traditional car when they did not have priority over the car. The need of being noticed or detected by an AV may not only be relevant for external road users: Passengers of automated shuttle services may want to be noticed by someone in an remote control room. In the study of Vlakveld et al. (2020), cyclists were more inclined to slow down in conflict situations at intersections with an AV rather than a traditional car approaching. In line with Parkin et al. (2022), we posit that trust represents a fundamental basic human need, which can be hierarchically organized at the bottom of the pyramid as shown by Figure 1. The need for trust can be translated into operationalizable indicators for all road users to be administered in questionnaires for the assessment of AVA ("I can trust the AV", "more attentive driver", "become complacent", "AV is reliable", "feel comfortable trusting life to beloved others", "fear loss of control", "being detected by AV") (see Table 1).

## 2.3.    Efficiency

Studies have revealed that efficiency, such as performance expectancy (or the perceived usefulness), and facilitating conditions (or the support of facilitating conditions supporting the use of AVs), influenced the behavioral intention to use AVs (Lehtonen et al., 2022; Nordhoff et al., 2020). The effect of perceived ease of use (i.e., effort expectancy) on the intention to use automated cars was ambiguous, with some studies reporting positive (Chen, Li, Gan, Fu, & Yuan, 2020; Madigan et al., 2016), or no effects (Benleulmi & Ramdani, 2022; Kettles & Van Belle, 2019; Madigan et al., 2017; Nordhoff et al., 2020). Reductions in travel time, travel costs, and fuel or energy consumption are other key aspects of efficiency, and key expected benefits of travelling with an AV (Merat & Lee, 2012; Szimba & Hartmann, 2020). These aspects may be particularly important for truck drivers, especially fleet owners, who perceive the AV as an opportunity to create a mobile workplace, promoting productivity by performing work-related tasks (Fröhlich et al., 2018). The acceptance of this new workplace by professional truck drivers is still unclear. Studies have shown that a large proportion of truck drivers is unaware of their built-in AV technology (Richardson, Doubek, Kuhn, & Stumpf, 2017). Another aspect of efficiency pertains to travel cost savings (e.g., fuel consumption and insurance costs), which mainly arise with a higher penetration rates of AVs (Xie & Liu, 2022). (Motor-)cyclists rated 'shorter travel times' unlikely to be a perceived benefit of AVs, and were undecided about the decrease in traffic congestion as a result of AVs (Pammer et al., 2021). As shown by Figure 1, we posit that efficiency follows safety and trust as basic human need. In other words, we propose the hypothesis that once manufacturers satisfy the need for safety and trust, users will strive for the satisfaction of the efficiency of AVs following the reasoning of Maslow and Lewis (1987). As fundamental need, efficiency can be translated into operationalizable indicators ("better driver", "more useful", "make travelling easier", "reach destination faster", "reduce travel time in congestion", "reduce travel costs", "help with parking / on (congested) motorways / in urban traffic", "better for the environment") (see Table 1).

## 2.4.    Comfort & pleasure

Comfort is another key factor impacting AVA (Peng et al., 2023). Peng et al. (2023) proposed in their conceptual framework that comfort is directly influenced by trust, and perceived safety. Motion comfort is a key aspect of comfort. Insufficient levels of motion comfort can lead to motion sickness (de Winkel, Irmak, Happee, & Shyrokau, 2023; Irmak, de Winkel, Pool, Bülthoff, & Happee, 2021), decrease in cognitive task performance, an increase in subjective workload, or discomfort. Ease of use, physical comfort, and engagement in secondary tasks were suggested as additional factors impacting comfort during automated driving (Peng et al., 2023). Other studies have revealed that respondents expressed an interest to use AVs while being impaired from alcohol, drug, medication use or tiredness (Cunningham, Regan, Horberry, Weeratunga, & Dixit, 2019; Lehtonen et al., 2022; Payre, Cestac, & Delhomme, 2014). This reflects our recent study in which drivers of partially automated cars reported to travel tired, impaired and in inclement weather conditions (Nordhoff et al., 2023). In the study of Lehtonen et al. (2022) travelling in darkness was a positive predictor of travelling more with an AV. Therefore, we postulate that 'travelling tired or impaired', 'travelling in inclement weather and visibility conditions' represents a need or preference of AV drivers. Similarly, it could be posited that 'travelling tired or impaired', and 'travelling in inclement weather and visibility conditions' also represents a need of external road users. External road users may be more prone to travelling tired or impaired or in inclement weather and visibility conditions with AVs on public roads given the programmed cautiousness of AVs (see Nordhoff et al., 2023). An online survey with truck drivers revealed that the expected driving pleasure was a primary motive for choosing the profession, with some truck drivers fearing the loss of driving pleasure due to automating the driving task (Richardson et al., 2017). We posit that the need for comfort represents a need that exists universally across user groups. It hierarchically follows the need for efficiency, which implies that the need for comfort will appear once the need for efficiency has been satisfied. The need for comfort and pleasure can be translated into measurable indicators for all road users ("arrive more comfortably", "more enjoyable", "driving tired or impaired", "using AV in adverse weather conditions", "use travel time for leisure activities", "use travel time for non-leisure activities", "reduce motion sickness"). The indicator "reduce motion sickness" is only applicable for AV passengers (see Table 1).

## 2.5.    Social influence

The role of social influence for technology adoption has been acknowledged by technology acceptance models (Ajzen, 1991; Venkatesh et al., 2012). Studies investigating AVA have shown that social influence did impact the behavioral intention to use AVs (Chen et al., 2020; Nordhoff et al., 2020; Zhang et al., 2019). We posit that the need for social appreciation from user's important social networks hierarchically follows the need for efficiency. It can be translated into operationalizable indicators for all road users ("People who are important to be would think that I should use an AV", "I would drive an expensive AV, because I can") (see Table 1).

## 2.6.    Well-being

The topic of mental health has entered the transportation arena, with the World Health Organization (WHO) acknowledging the role of mobility for the prevention and treatment of mental disorders (M. Conceição et al., 2022; WHO, 2019). Mental health is defined as a state of well-being, enabling the realization of one's own abilities, coping with the normal stresses of life, working productively and fruitfully, and contributing to the community (WHO, 2004). Mental health has been commonly measured by affective states (emotions and mood), well-being and satisfaction with life or travel, stress and mental health disorders (M. A. Conceição et al., 2023). Mobility also has an impact on other dimensions of mental well-being, such as social inclusion, stress, workload, driving anxiety, or even mental disorders such as depression (M. Conceição et al., 2022). AVs can have a positive impact on the mental and physical well-being of drivers: Automated passenger vehicles can directly reduce mental workload, stress, and aggressive driving, making driving more relaxing and increasing drivers' situational awareness as drivers are no longer required to perform most of the tactical and operational parts of driving (Nordhoff et al., 2023). Conversely, AVs can have a negative impact on its passengers: AV drivers disengaged the automation as a result of passengers' discomfort and lack of trust in the system due to the automation's erratic, harsh, and unpredictable behavior (Nordhoff & De Winter, under review). This study also revealed that other road users interacting with the partially automated vehicles (i.e., Tesla Autopilot, FSD Beta) were confused and angry at the behavior of the automation. Scientific studies provide scientific evidence for the mental health (e.g., loneliness, depression, and anxiety), and physical health issues (e.g., back disorders, heart disease, obesity) of professional drivers due to the demanding and irregular work schedules, and the difficulty to maintain a healthy lifestyle (Dahl et al., 2009; Ji-Hyland & Allen, 2022; Sousa & Ramos, 2018). We posit that well-being hierarchically follows the need for social influence, and arises when the lower-level needs are satisfied. It can be translated into operationalizable indicators for all road users ("better awareness of surroundings", "make driving less stressful"; "make driving more relaxing", "arrive less tired", "reduce aggression on the road") (see Table 1).

## 3.    AVA pyramid

The present study organizes the AVA road user needs and preferences hierarchically in the form of a pyramid as shown in Figure 1. The AVA pyramid displays user needs and preferences ordered from basic, fundamental needs at the bottom to higher-level user needs and preferences at the top of the pyramid. The pyramid assumes that higher-level needs (i.e., needs at higher levels of the pyramid) arise with the satisfaction of the needs at lower levels of the pyramid.

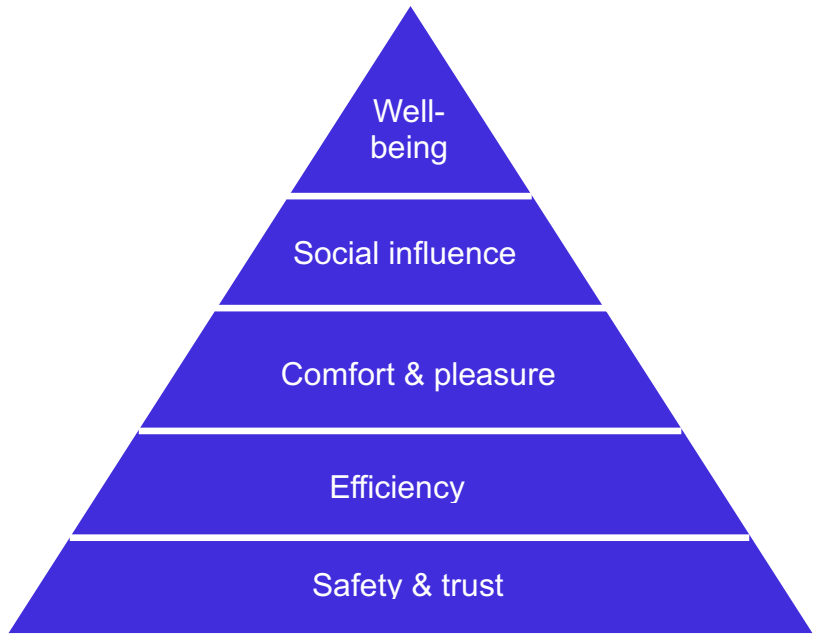

*Figure 1.* AVA pyramid displaying road user needs and preferences ordered from basic fundamental needs at the bottom to higher-level user needs and preferences at the top of the pyramid.

These needs and preferences are translated into operationalizable indicators as shown by Table 1, which shows the applicability of these indicators per road users.

*Table 1.* Overview of general indicators and relevance per road user

| Need | Indicator | Road users | | | | | |
|---|---|---|---|---|---|---|---|
| | | **Drivers** | **Passengers** | **Truck drivers** | **Other drivers** | **Pedestrians** | **(Motor-) cyclists** |
| **Safety** | Arrive safer | X | X | X | X | X | X |
| | Feel safer | X | X | X | X | X | X |
| **Trust** | Trust the AV | X | X | X | X | X | X |
| | More attentive | X | X | X | X | X | X |
| | Become complacent | X | X | X | X | X | X |
| | AV is reliable | X | X | X | X | X | X |
| | Feel comfortable trusting life of loves ones to AV | X | X | *NA* | X | X | X |
| | Fear loss of control | X | X | X | X | X | X |
| | Being detected by AV | *NA* | *NA* | X | X | X | X |
| **Efficiency** | Better driver | X | X | X | X | X | X |
| | More useful | X | X | X | X | X | X |
| | Make travelling easier | X | X | X | X | X | X |
| | Reach destination faster | X | X | X | X | X | X |
| | Cope with congestion | X | X | X | X | *NA* | X |
| | Reduce travel costs | X | X | X | X | *NA* | X |
| | Help with parking / on (congested) motorways / in urban traffic | X | *NA* | X | X | *NA* | *NA* |
| | Better for the environment | X | X | X | X | *NA* | X |
| **Comfort & pleasure** | Arrive more comfortably | X | X | X | X | X | X |
| | More enjoyable | X | X | X | X | X | X |
| | Driving tired or impaired | X | X | X | X | X | X |
| | Using AV in adverse weather conditions | X | X | X | X | X | X |
| | Use travel time for leisure activities | X | X | X | X | X | X |
| | Use travel time for non-leisure activities | X | X | X | X | X | X |
| | Reduce motion sickness | *NA* | X | *NA* | *NA* | *NA* | *NA* |
| **Social influence** | Social influence | X | X | *NA* | X | X | X |
| **Well-being** | Better awareness of surroundings | X | X | X | X | X | X |
| | Make driving less stressful | X | X | X | X | X | X |
| | Make driving more relaxing | X | X | X | X | X | X |
| | Arrive less tired | X | X | X | X | X | X |
| | Reduce aggression on the road | X | X | X | X | X | X |
| **Acceptance** | Shift from train or airplane to car on longer trips | *NA* | *NA* | X | X | X | X |
| | Plan to use | X | X | X | X | X | X |
| | Intend to use | X | X | X | X | X | X |
| | Buy AV as next car | X | X | X | X | *NA* | *NA* |
| | Make more daily trips with AV | X | X | *NA* | *NA* | *NA* | *NA* |
| | Make more long-distance trips with AV | X | X | *NA* | *NA* | *NA* | *NA* |

| | Travel less by public transport | X | X | *NA* | X | X | X |
|---|---|---|---|---|---|---|---|
| | Travel less when AVs are around | *NA* | *NA* | *NA* | X | X | X |

## 4. The road ahead

This study proposed a new methodological approach for the assessment of AVA from the perspective of road users inside and outside AVs. A pyramid was developed, which provides a hierarchical organization of user needs. The pyramid posits that safety and trust in AVs represent basic and fundamental needs at the bottom of the pyramid. The need for safety and trust exists universally across road users inside and outside of AVs. After the need for safety and trust is fulfilled, the need for efficiency arises as the third-lowest layer of the pyramid. After the satisfaction of the need for efficiency, users may want to strive for the realization of the need for comfort and pleasure as the fourth-lowest layer of the pyramid. Social influence – the social appreciation of the use of AVs in the networks of AV users – represents the need that users strive to achieve after their fulfilment of their need for comfort and pleasure. At the top of the pyramid is the need for a user's well-being, which represents the highest-level need of AV users, and reflects how AV users feel in their interaction with AVs.

This paper translated these needs into operationalizable indicators or questionnaire items that can be administered in studies for the assessment of AVA pre- and post- AV experience. The questionnaire captures the most important needs of road users. We recommend future studies to validate the questionnaire, and contribute to the development of an efficient and standardized metric for the assessment of AVA.

## 6.    Acknowledgements

The authors would like to thank all partners within the Hi-Drive project for their cooperation and valuable contribution. This project has received funding from the European Union's Horizon 2020 research and innovation programme under grant agreement No 101006664. The role responsibility of this publication lies with the authors. Neither the European Commission nor CINEA – in its capacity of Granting Authority – can be made responsible for any use that may be made of the information this document contains.